School of Automation, Central South University, China

# Optimal Placement of Capacitor in Distribution System Using Particle Swarm Optimization

Izhar Ul Haq
*School of Automation*
*Central South University*
Hunan, China
izharctl9@gmail.com

Asifa Yousaf
*School of Automation*
*Central South University*
Hunan, China
asifayusuf76@gmail.com

Javeria Noor
*School of Automation*
*Central South University*
Hunan, China
javerianoor19@gmail.com

Muhammad Shams ur Rehman
*School of Automation*
*Central South University*
Hunan, China
shams.ee@yahoo.com

*Abstract*—In power systems, the incorporation of capacitors offers a wide range of established advantages. These benefits encompass the enhancement of the system's power factor, optimization of voltage profiles, increased capacity for current flow through cables and transformers, and the mitigation of losses attributed to the compensation of reactive power components. Different techniques have been applied to enhance the performance of the distribution system by reducing line losses. This paper focuses on reducing line losses through the optimal placement and sizing of capacitors. Optimal capacitor placement is analyzed using load flow analysis with the Newton-Raphson method. The placement of capacitor optimization is related to the sensitivity of the buses, which depends on the loss sensitivity factor. The optimal capacitor size is determined using Particle Swarm Optimization (PSO). The analysis is conducted using the IEEE-14 bus system in MATLAB. The results reveal that placing capacitors at the most sensitive bus locations leads to a significant reduction in line losses. Additionally, the optimal capacitor size has a substantial impact on improving the voltage profile and the power loss is reduced by 21.02% through the proposed method.

I. INTRODUCTION

As distributed systems extend day by day being away from the generation station results in different losses which results in poor power factor and less useful power. Instability occurs in a power system when its voltages unmanageably change between load (Power demand) and generation (Power generation), due to outdated equipment, failure to control voltage and outage of controlled mechanism in the system. This type of issue i.e. instability occurs due to different sorts of undesirable components of power inclusion in the power system.

The losses due to the reactive power in the system are minimized by using shunt capacitors, FACT devices and distributed generators (DGs) installed in the distribution system [1]. Power loss reduction in the network is a major concern for electrical Engineers through the contribution of effective electric distribution utilities Hence, to obtain lower power losses and an improved voltage profile optimum size of the capacitor must be installed at the optimum location. Optimum placement and size of the capacitor were familiarly experimented over decades for better result.

There are different techniques for optimization purposes such as numerical programming, heuristics and artificial intelligence. A heuristic method is proposed in reference [2]. The conjunction of analytical with heuristics methods for the placement of capacitors was first presented by Neagle [3], and then by Cook [4]. Load flow based on a heuristic algorithm is presented for determination and outlines with a small amount of power losses. In reference [5] and [6], the problem of capacitor placement is solved through nonlinear integer programming with a decomposition method and a power flow algorithm. Reference [7], solved this issue using a programming method considering continuous variables for the location and size of capacitors. The genetic algorithm (GA) based method was proposed in [8], further increased and contributed in reference [9], and then after other researchers with more functions. Reference [10], solved this with two stages in GA. The result obtained in the first stage is further made much better using sensitivity-based heuristics in the second stage. Reference [11], solved many problems using GA in feeders which consist of annual energy loss reduction, system capacity release, and load reduction. Nowadays, many evolutionary methods have been proposed such as the stochastic method, GA method, evolutionary programming and PSO. In other papers the writer used Tabu search heuristic; others through micro-genetic algorithms; some relied on a hybrid genetic algorithm. In [12], the author has used the PSO for the optimization of losses using the loss sensitivity factor by the Gauss-Seidel method and has obtained a 14.25% power loss reduction. The article [13], has used Newton Raphson method for the allocation of DG's in IEEE-33 and 69 bus systems.

Finding the sensitivity of different buses and sizing the capacitor is the main objective of this paper and we have used the Newton-Raphson method. In this paper, we have used Particle Swarm Optimization (PSO) to find the size and placement of capacitors in the IEEE-14 bus system for power

---

loss reduction and improvement of power factor under MATLAB coding. PSO over other methods has many more benefits specified in [14]. Power flow equations are employed to calculate the power distribution across the buses in the IEEE-14 bus system. This power flow analysis serves as the basis for determining the sensitivity of the load buses, which is accomplished using the Newton-Raphson method. Consequently, this approach facilitates the identification of optimal capacitor placement, ensuring the minimization of power loss for enhanced system performance.

## II. PROBLEM STATEMENT

The problem intended is to determine the location and size of the capacitor optimally in the IEEE-14 bus system. The objective of this paper is to minimize the total annual cost of capacitors and reduction of power losses, which is given by

$$\text{Cost} = K^P P_{loss} + \sum_{j=1}^{j} K_j^c Q_j^c \quad (1)$$

where $P_{loss}$ is the total power losses, $K^p$ is the annual cost per unit of power losses, $K^c$ is the capacitor annual cost, $Q_j^c$ is the shunt capacitor size placed at bus $i$, and $j$ is the number of capacitors in the respective bus.

## III. CONSTRAINTS

The following constraints and parameters that should be fulfilled during the iteration to find the losses are described.

### A. Shunt capacitor limits

$$Q_{max}^c \leq Q_{total} \quad (2)$$

where, $Q_{max}^c$ maximum capacitor size allowed and $Q_{total}$ is the total reactive load attached to a bus.

### B. Bus bar voltage

The specific constraint for the bus bar voltage is given below;

$$V_{min} < V_i < V_{max} \quad (3)$$

### C. Power Flow limits

The power flow limits in a specific bus are limited to;

$$\text{Flow}_k < \text{Flow}_k^{max} w \quad (4)$$

where, $\text{Flow}_k$ is the power flow in the kth line and $\text{Flow}_k^{max}$ is the highest power that can pass through the specific bus.

## IV. SENSITIVITY ANALYSIS

It is a way to find the most sensitive bus (buses which have maximum variation with the change in the load). The main purpose of capacitor installation is to reduce line losses. This can be done by placing the capacitor at the most favourable place which will have maximum contribution toward loss reduction. This will reduce the search space for particle swarm optimization. Sensitive buses are found by power flow through Newton Raphson method [15]. All the active and reactive power losses are losses found by power flow. So, we will have a better knowledge about the sensitive bus.

$$P_i = \sum_{j=1}^{N} |V_i||V_j| \begin{pmatrix} G_{i,j} \cos(\delta_i - \delta_j) + \\ B_{i,j} \sin(\delta_i - \delta_j) \end{pmatrix} \quad (5)$$

$$Q_i = \sum_{j=1}^{N} |V_i||V_j| \begin{pmatrix} G_{i,j} \sin(\delta_i - \delta_j) + \\ B_{i,j} \cos(\delta_i - \delta_j) \end{pmatrix} \quad (6)$$

So, both the equations are non-linear, so we have used an iterative process to solve it and its graph is given in Fig.1.

## V. IMPLEMENTATION OF NEWTON RAPHSON METHOD

First of all, read the 14-bus system data and specified load and generator also read the voltage specification it regulated bus after reading the overall data of 14-bus system then form admittance matrices of 14 by 14 order. While the Newton-Raphson method is a numerical method it needs some initial guess value to initialize the voltage and angle of each bus in the 14-bus system. Then start the iteration cycle and updates the voltage and angle of each bus to satisfy the specified condition of load and generation, and after updating the voltage and angle of 14 buses check the specified condition if it specifies the condition then stop the iteration and the result will be the voltage and angle of each bus in 14 bus system if it not satisfied then update it voltage and angle until it satisfied. After stopping iteration check the result of which bus gives more losses than others and consider that bus a sensitive.

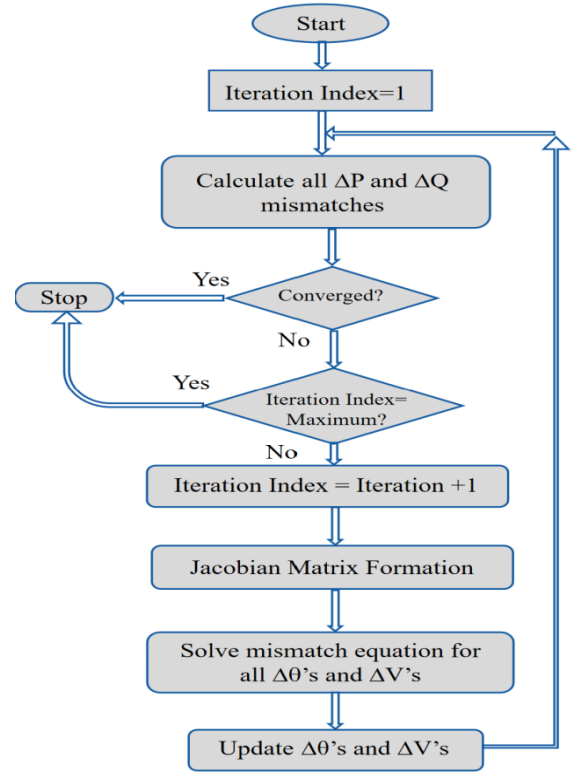

Fig. 1. Flow chart for sensitivity analysis

### A. Steps For Newton Raphson

1. Initialize the inductance Matrix YBUS
2. Consider initial voltages to be as follows:

$$V_i = V_{(spec)}^{(<0^0)} \quad (7)$$

   for all PV buses.

3. At (r+1) iteration, calculate $P_i^{r+1}$ at all the V and PQ buses and $Q_i^{r+1}$ at all the PQ bus, using values from the previous iteration, $V_i$. The formulas to be used are:

$$V_i = 1 > 0^0 \quad (8)$$

$$P_i = \sum_{j=1}^{N} |V_i||V_j| \begin{pmatrix} G_{i,j} \cos(\delta_i - \delta_j) + \\ B_{i,j} \sin(\delta_i - \delta_j) \end{pmatrix} \quad (9)$$

$$Q_i = \sum_{j=1}^{N}|V_i||V_j|\begin{pmatrix} G_{i,j}\sin(\delta_i-\delta_j) - \\ B_{i,j}\cos(\delta_i-\delta_j) \end{pmatrix} \quad (10)$$

4. Calculate the Power mismatches (Power Residues)

$$\Delta P_i^r = P_{i(spec)} - \Delta P_i^{r+1} \quad (11)$$
$$Q_i^r = Q_{i(spec)} - \Delta Q_i^{r+1} \quad (12)$$

for all PV and PQ bus.

5. Compute the Jacobian Matrix [J(r)] using V(r) and its element spread over sub-matrices.
6. Compute

$$\begin{bmatrix}\Delta\delta(i)^r \\ \Delta V(i)^r\end{bmatrix} = [[J]^{-r}]\begin{bmatrix}\Delta\delta(i)^r \\ \Delta V(i)^r\end{bmatrix} \quad (13)$$

7. Update the variable as follows

$$\delta_i^{r+1} = \delta_i^r + \Delta_i \quad (14)$$
$$V_i^{r+1} = V_i^r + \Delta V_i \quad (15)$$

8. Go back to step 3 and iterate until the power mismatch with required tolerance.

## VI. PARTICLE SWARM OPTIMIZATION TECHNIQUE

Dr. Eberhart and Dr. Kennedy created the particle swarm optimization (PSO) method in [16], a population-based evolutionary computation technique modeled by the social behavior of fish schools and flocks of birds. It makes use of the particle population with a specific speed that can go into hyperspace. Particle speeds settle with each repetition, and the optimal collection of historical locations for every particle, together with its surrounding particles, are preserved. The ideal location and dimensions based on the customized fitness function is discussed in details [17, 18].To improve the voltage profile and reduce the active power loss, it presents a new approach to determine the position and optimum dimension of the capacitor in the radial distribution system. The placement and size of the capacitors are made by the susceptibility and optimization factors of the particles respectively, against the drag [19]. It emphasizes the basic features and advantages of the PSO compared to various other optimization algorithms and also discusses possible uses of the PSO in the field of the electric energy system and possible theoretical studies.

Two new objective functions are identified by adding value to reliability, loss costs and investment costs. This problem is solved by using an algorithm based on the optimization of particle droplets in the distribution network [20]. This approach efficiently searches the solution region and can handle continuous state variables with ease. Nevertheless, discrete and continuous variables can also be handled by expanding this strategy. PSO is an algorithm for evolutionary computation. Each particle in the PSO enters the search field at a speed dynamically adjusted according to its own and its neighboring particle flight experiences. At any moment, the particle speed changes to the best individual *pbest* and global best *gbest* positions. Particle acceleration is a random term and different random numbers are generated for the positions of the *pbest* and *gbest* particles.

Generally, any optimization problem can be defined as

$$Min\ (F(x))$$
$$S_i^{k+1} = \omega * S_i^k + c_1 * rand(p_{best} - P_i^k) + c_2 * rand(g_{best} - P_i^k) \quad (16)$$

where $c_1, c_2 = 2, S_i$ is the velocity of the particles. The best position attained by any particle is recorded and represented as:

$$P_i^{k+1} = P_i^k + S_i^{K+1} \quad (17)$$
$$pbest_i = (pbest_{i,1}, pbest_{i,2}\ \ldots..\ pbest_{i,d}) \quad (18)$$

The best position achieved by a particle among the population is represented by *gbest*. The change in velocity for a particle can be represented as:

$$S_i = (S_{i,1}, S_{i,2}\ \ldots..\ S_{i,d}) \quad (19)$$

The new speed and the actual speed of every particle can be calculated using the *pbest* and *gbest* . The current velocity and next iteration velocities can be expressed in formulas

$$S_{min} \leq S_i \leq S_{max} \quad (20)$$

TThe resolution or fitness with which regions between the current position and the target position are examined is determined by the parameter $S_{max}$. in equation (20). If $S_{max}$ is too small, particles may not explore sufficiently beyond the local solutions [21].

Constants *c*1 and *c*2 represent the stochastic acceleration and its value is taken as $c1 = c2 = 2$, $\omega$ is the inertia of the particles. Suitable selection helps balancing between local and global exploration and exploitation and results in fewer iterations required for sufficiently optimal solutions. Generally taken between 0.9 and 0.4.

The schematic for PSO is given in Fig. 2.

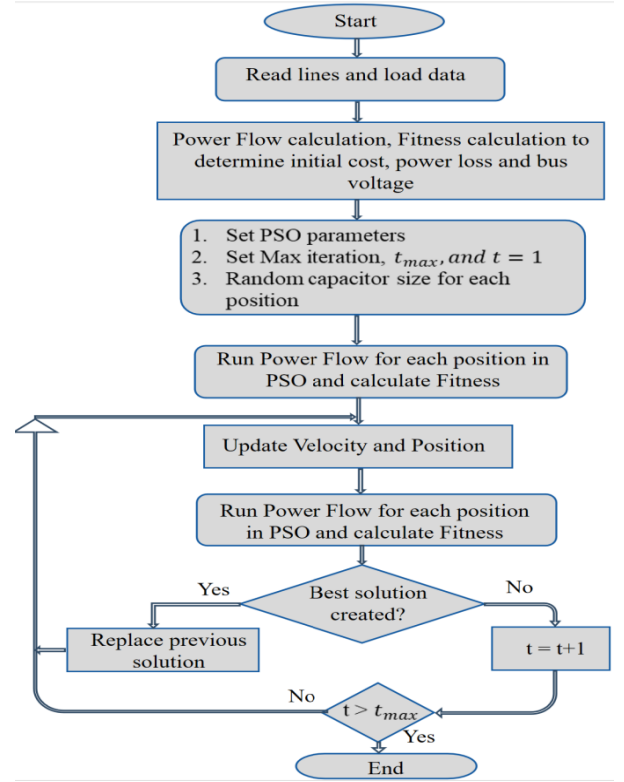

Fig. 2. Flow chart for PSO implementation

## VII. IMPLEMENTATION OF ALGORITHM

### A. Selection of candidate bus

1) Input of the IEEE 14 bus system impedances real

and reactive power data.
2) Sensitivity Analysis.
3) Buses with higher Sensitivity is taken as a candidate bus.

B. Optimization using Particle Swarm algorithm

1) (PSO) control data
2) Initialization of Particle Population
3) Compute capacitor cost
4) Find the total cost function (fitness function)
5) Record the previous best performance for each particle and save as pbest
6) Best performance for all the particles is recorded as gbest.
7) Find the velocity and position of each particle
8) Update particle variables.
9) If the obtained value of pbest is better than the previous than replace it with new
10) Calculate best values of all pbest and save as gbest
11) Record the resultant solution

## VIII. IEEE-14 BUS SYSTEM

IEEE-14 bus is among the Power system model that's usually used for research purposes among many researchers because of its constant resistance, reactance and susceptance with fixed buses, generators and transformers installed. For each of the buses in 14-bus system, four variables that characterize the buses electrical conditions. The four variables are real and reactive power injection $P_k$ and $Q_k$ respectively, and voltage magnitude $V_k$ and angle $\emptyset_k$, where k is any bus number. Based on these variables, buses are categorized into three big categories.

A. PV Buses (voltage-controlled buses)

There are two variables known, that are $P_k$ and $V_k$ but the other two, $Q_k$ and $\emptyset_k$ are unknown.

B. PQ Buses (load buses)

There are two variables $P_k$ and $Q_k$ are known but $V_k$ and $\emptyset_k$, are unknown variables. All load buses fall into this category.

C. Swing Bus

In the swing bus, only the generator is attached but not the load. The other two terms used for this bus and swing bus and reference bus. The swing bus may be any bus having an attached generator. For swing buses $V_k$ and $\emptyset_k$ are known and the other variables are unknown. The detailed schematic of the IEEE-14 system is given in Fig.2.

## IX. SIMULATION AND RESULTS

The proposed algorithm has been tested on the IEEE-14 bus system. The annual cost per unit of power loss $K_p$ is taken as 168 $/KW. The voltage limits are taken as $V_{max} = 1.06$ KV and $V_{min} = 0.95$ KV. After running the power flow without any capacitor compensation, the total power losses were 25.74kW and the annual cost function of the power losses was 4325.233$. Loss-sensitive factors are calculated from the load flow for each line.

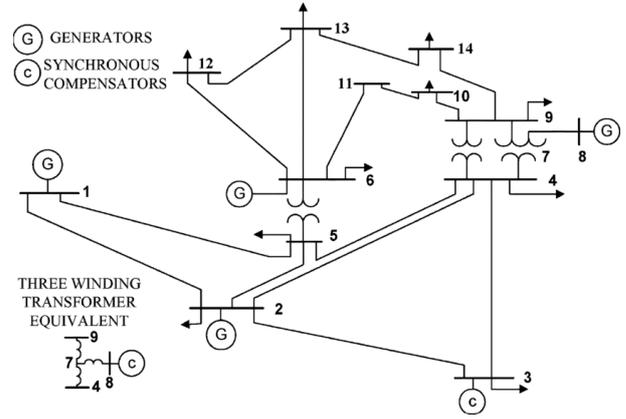

Fig. 3. Schematic of IEEE-14 Bus system

The buses requiring compensation are given as 6 and 9. The rest of the buses are healthy and do not need any compensation. The sizing of the capacitor is done by the PSO algorithm. After performing 40 iterations the results obtained are shown in the Table.1 including the maximum and minimum total cost and active power losses.

TABLE I. IEEE-14 BUS SYSTEM BEFORE AND AFTER COMPENSATION

| | | |
|---|---|---|
| Maximum KW (Worst) | Before Compensation | 25.7454 kW |
| Minimum KW (Best) | After Compensation | 14.2946 kW |
| Maximum $/Year (Worst) | Before Compensation | 4325.2337$ |
| Minimum $/Year (Best) | After Compensation | 2401.4866 $ |

The total Power loss for the whole process is shown in Fig.4 concerning the number of iterations. The change becomes very minute after the *4th* iteration and becomes constant after the *40th* iteration. minimum total cost, active power losses.

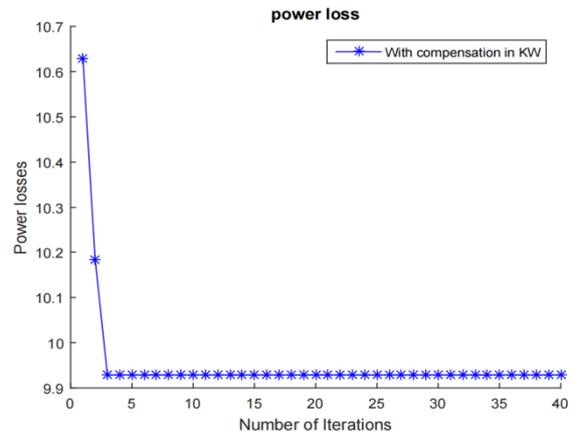

Fig. 4. Power loss profile

The voltage profile before and after the compensation for all 14 buses is shown in Fig.5, with minimum voltage on the 14th bus and maximum on the 2nd bus. Using the optimal placement of the capacitor in the optimal position of the IEEE-14 bus system has improved the voltage profile of every bus.

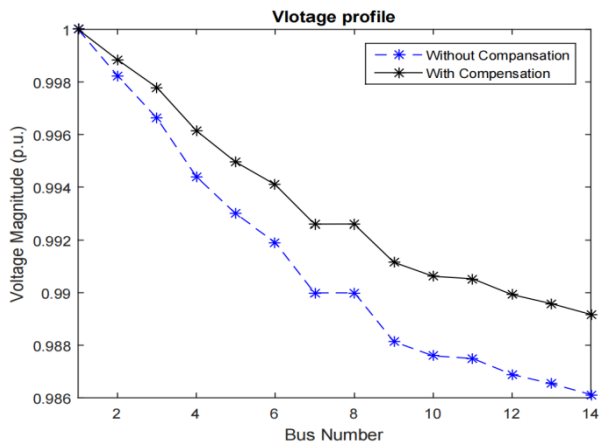

Fig. 5. Voltage profile before and after compensation

The Loss sensitivity factor using Newton Raphson method is given in the Table 2. The most sensitive bus so far obtained from this method is 9th bus and then after the 6th bus in the system. The capacitor size obtained by using the particle swarm optimization is placed at the required location to obtain power loss reduction and improvement of voltage profile. The values for the optimum size and respective voltage improvement are given in Table 3.

TABLE II. LOSS SENSITIVITY FACTOR

| From (Bus) | To (Bus) | LSF | LSF in Dest | End Bus | Bus Voltage | Norm(i) |
|---|---|---|---|---|---|---|
| 1 | 2 | 0.293874 | 0.093722 | 2 | 1.003 | 1.01445 |
| 1 | 5 | 0.047487 | 0.099031 | 5 | 0.993 | 1.04353 |
| 2 | 3 | 0.066776 | 0.021234 | 3 | 0.997 | 1.05676 |
| 2 | 4 | 0.032567 | 0.013471 | 4 | 0.996 | 1.10860 |
| 2 | 5 | 0.016772 | 0.082655 | 5 | 0.984 | 1.09673 |
| 3 | 4 | 0.018994 | 0.045763 | 4 | 0.983 | 1.03522 |
| 4 | 5 | 0.064771 | 0.028976 | 5 | 0.992 | 1.09524 |
| 6 | 11 | 0.004234 | 0.009963 | 11 | 0.982 | 0.99645 |
| 6 | 12 | 0.067882 | 0.010789 | 12 | 0.987 | 1.00356 |
| 6 | 13 | 0.136643 | 0.037781 | 13 | 0.967 | 1.02389 |
| 7 | 8 | 0.073833 | 0.009762 | 8 | 0.985 | 0.97352 |
| 7 | 9 | 0.012239 | 0.007872 | 9 | 0.986 | 0.93456 |
| 9 | 10 | 0.099832 | 0.056789 | 10 | 0.989 | 1.03546 |
| 9 | 14 | 0.000128 | 0.035637 | 14 | 0.9873 | 1.05673 |
| 10 | 11 | 0.027997 | 0.008892 | 11 | 0.990 | 1.10547 |
| 12 | 13 | 0.05749 | 0.003558 | 13 | 0.95 | 1.03893 |
| 13 | 14 | 0.67834 | 0.010356 | 14 | 0.989 | 1.03567 |

TABLE III.

| Bus No | With Out Capacitor | | Proposed PSO Capacitor | |
|---|---|---|---|---|
| | Size(MVAr) | V(pu) | Size(MVAr) | V(pu) |
| 6 | 0 | 0.993 | 4.05 | 0.9901 |
| 8 | 0 | 0.982 | 4.10 | 0.9941 |
| 9 | 0 | 0.986 | 0.67 | 0.9939 |
| Total (MVar) | 0 | --- | 8.88 | --- |
| Active power loss | 25.74 KW | | 14.29 KW | |

## X. CONCLUSION

This research uses particle swarm optimization to solve the optimal capacitor placement and sizing problem in an IEEE-14 bus system. One of the optimizations method is particle swarm optimization, which is used to optimize the size of capacitor. In contrast with other conventional approaches, PSO is a population-based methodology. As a result, PSO with Newton-Raphson method is less likely trapped in local minima. The optimal position is reached because the quality of the solution is independent of the original population's starting point in the search space. Tested on an IEEE-14 bus system, the suggested technique produces far better results than smaller-iteration algorithms documented in the literature. The benefits of the problem solution include an enhanced voltage profile and power factor. the reduced flow over the distribution, transmission, and transformer lines as well as the decrease in losses as a result of reactive power compensation. It is possible to raise system load without adding more cables by reducing the flow through the wires.

ACKNOWLEDGMENT

Not applicable.